\documentclass[twocolumn]{aastex701}

\usepackage{bm,booktabs,enumitem} 
\usepackage{newtxmath,newtxtext}
\usepackage{xcolor}
\usepackage{subcaption}

\begin{document}

\title{Charge Exchange Illuminates Turbulent Cold–Hot Interfaces}

\author[0009-0003-4557-1106]{Yizhan Zhang} 
\affiliation{Center for Astronomy and Astrophysics and Department of Physics, Fudan University, Shanghai 200438, P.R.China.}
\email{yizhanzhang23@m.fudan.edu.cn}

\author[0000-0001-9658-0588]{Suoqing Ji}
\affiliation{Center for Astronomy and Astrophysics and Department of Physics, Fudan University, Shanghai 200438, P.R.China.}
\affiliation{Key Laboratory of Nuclear Physics and Ion-Beam Application (MOE), Fudan University, Shanghai 200433, P.R.China.}
\email[show]{sqji@fudan.edu.cn}

\author[0000-0003-1628-5058]{Guiyun Liang}
\affiliation{National Astronomical Observatories, Chinese Academy of Sciences, Beijing 100101, P.R.China.}
\email{gyliang@bao.ac.cn}

\author[0000-0002-8329-6606]{Shuinai Zhang}
\affiliation{Key Laboratory of Dark Matter and Space Astronomy, Purple Mountain Observatory, Chinese Academy of Sciences, Nanjing 210023, P.R.China.}
\email{snzhang@pmo.ac.cn}

\author[0000-0002-2941-646X]{Zhijie Qu}
\affiliation{Department of Astronomy, Tsinghua University, Beijing 100084, P.R.China.}
\email{quzhijie@tsinghua.edu.cn}

\begin{abstract}

Cold-hot interfaces are ubiquitous in the multiphase circumgalactic medium (CGM), galactic winds, and the Galactic Center (GC), yet their microscopic contact regions are difficult to observe. Charge exchange (CX), in which a highly ionized metal ion captures an electron from a neutral donor and emits soft X-ray lines, directly probes these interfaces. We develop a subgrid CX model coupled to a nonequilibrium ionization (NEI) network in hydrodynamic simulations of radiative turbulent mixing layers. CX soft X-rays are confined to a thin, highly corrugated cold-hot interface (corrugation factor $\sim10$), whereas thermal emission fills the hot phase. Turbulent transport sustains high-charge ions at $T\sim10^4$--$10^5$~K, where they encounter neutral donors. The low-temperature \ion{O}{8} fraction exceeds that in a uniformly cooling isobaric parcel, demonstrating the importance of mixing to the ionization history. CX depletes this cool ion population, while continued entrainment maintains the supply. Including a metagalactic ionizing background changes the ion distribution and enhances CX surface brightness by a factor of a few in our CGM comparison, but preserves the cool origin of most CX emission. In the limit of sufficient neutral column, CX surface brightness is ion-supply limited: higher donor density increases local emissivity but shortens the reaction layer. The system-wide CX-to-thermal soft X-ray ratio also depends on the volume-to-interface scale $V/A_\mathrm{int}$, making CX a potential probe of multiphase structure in the CGM, galactic winds, and the GC.

\end{abstract}


\section{Introduction} \label{sec:intro}

Multiphase gas is ubiquitous in astrophysical systems, from the circumgalactic medium (CGM) and galactic winds to the Galactic Center (GC). Across these environments, cold neutral and molecular gas coexists with warm ionized material and hot X-ray-emitting gas \citep{tumlinson2017circumgalactic,thompson2024theory,ponti2015xmm}. Their evolution is governed not only by global accretion, gravity, and feedback, but also by the exchange of mass, momentum, and energy across cold-hot interfaces. In the CGM, this exchange helps regulate gas accretion, metal transport, and baryon recycling between galaxies and the intergalactic medium \citep{tumlinson2017circumgalactic,faucher2023key}; in the GC, it occurs amid dense molecular gas, massive star formation, supernova feedback, and nuclear activity \citep{ponti2015xmm,kruijssen2014controls,heywood2019inflation}. Recent HST and JWST observations of M82 provide a particularly vivid view of such coupling in a galactic wind, resolving H$\alpha$-emitting clouds and PAH-bearing cool filaments within the hot outflow and showing that dust-bearing material persists several kiloparsecs above the disk \citep{lopez2025observational,lopez2026jwst,cronin2026jwst}. Despite their diverse range in density and physical scale, these systems are therefore linked by the same local question: how does cold gas interact with the surrounding hot gas?

Radiative turbulent mixing layers provide a theoretical framework for this cold-hot interface physics. Shear at an interface drives the Kelvin-Helmholtz (KH) instability and turbulence, continuously exchanging mass, momentum, and energy between the two phases \citep{begelman1990turbulent,slavin1993turbulent}. Recent progress in the theory and simulation of radiative mixing layers emphasizes the importance of radiative cooling in this process: mixing first produces intermediate-temperature gas, typically around $10^5$~K, near the peak of the cooling curve where cooling is much more efficient than either the hot or cold phase; this gas can then cool and condense back onto the cold phase, allowing cold gas to survive or grow \citep{kwak2010numerical,ji2019simulations,fielding2020multiphase,tan2021radiative,sharma2025universal}. The fate of cold gas depends on the Mach number of the velocity shear, cooling strength, and geometry of the interface \citep{tan2021radiative,yang2023radiative,chen2026from,chen2026clumps}.

Despite the crucial role of cold-hot interfaces in regulating the evolution of multiphase gas in the CGM, galactic winds, and the GC, they remain difficult to probe observationally. A clue comes from a nearby ``multiphase'' system in our own Solar System: cometary X-ray emission, produced by charge exchange (CX) between the hot, highly ionized solar wind and cold cometary gas \citep{lisse1996discovery,cravens1997comet}. In a CX reaction, a highly ionized metal ion captures an electron from a neutral atom or molecule; the captured electron then cascades radiatively and emits characteristic soft X-ray lines. The process therefore selects the microscopic contact region where hot ions and cool neutral gas interpenetrate, providing a direct probe of cold-hot interfaces. It also produces line ratios that differ from ordinary collisionally ionized plasma, including enhanced forbidden lines in He-like triplets and strong high-$n$ cascades. CX-induced X-ray emission was recognized early on as a potentially significant contributor to diffuse X-ray backgrounds from the interstellar and intracluster medium \citep{lallement2004contribution,gu2018charge}. Observations of M82 and nearby star-forming galaxies suggest that CX might contribute substantially to soft X-ray line emission from galactic outflows \citep{liu2011m82,zhang2014m82,wu2020outflows}, and XMM-Newton spectroscopy of M51 indicates that CX might be a major component of diffuse disk emission \citep{zhang2022m51}.

Despite the potentially significant contribution of CX to soft X-ray emission and its unique diagnostic power for cold-hot interfaces, a quantitative theory of CX coupled to hydrodynamics remains largely unestablished, and attempts to incorporate it self-consistently into astrophysical hydrodynamic simulations are sparse. On the numerical side, the difficulty is intrinsically subgrid: a single-fluid simulation does not separately evolve ion and neutral velocity distributions, while the CX rate depends on ionization state, neutral abundance, relative velocity between ions and neutrals, and velocity-dependent atomic cross sections. On the theoretical side, thanks to recent progress on the physics of radiative turbulent mixing layers, we now have a better understanding of the structure and evolution of cold-hot interfaces, which provides a foundation for modeling and comprehending CX physics in these environments.

In this paper, we develop the first subgrid model for CX in hydrodynamic simulations and apply it to radiative turbulent mixing layers representative of the CGM and GC. The paper is organized as follows. \S\ref{sec:cx_model} describes the CX subgrid model for hydrodynamic implementation.
\S\ref{sec:methods} presents the numerical methods and simulation setup. \S\ref{sec:results} presents the simulation results. We finally discuss the implications of our findings and conclude in \S\ref{sec:discussion_conclusions}.

\section{Subgrid Model for Charge Exchange} \label{sec:cx_model}

In this section, we describe our subgrid model for CX in hydrodynamic simulations. The model is designed to be compatible with single-fluid simulations that do not separately evolve ion and neutral velocity distributions. The model also computes the resulting soft X-ray emission from CX reactions.

\subsection{Nonequilibrium Ionization with Charge Exchange} \label{subsec:nei}

The metal ion fractions are evolved with a nonequilibrium ionization (NEI) network by solving the following set of coupled ordinary differential equations:

\begin{align}
    \frac{\partial n_i}{\partial t} &= n_e \bigl[ n_{i+1} \alpha_{i+1} + n_{i-1} C_{i-1} - n_i (\alpha_i + C_i) \bigr] \notag \\
    & + n_{i-1}\Gamma_{i-1} - n_i\Gamma_i + X_{i+1} - X_i,
    \label{eq:nei}
\end{align}
where $n_i$ is the number density of the $i$-th ion, $n_e$ is the electron number density, $\alpha_i$ denotes the total recombination coefficient (including radiative and dielectronic recombination) from ion $i$ to ion $i-1$, $C_i$ represents the collisional ionization coefficient from ion $i$ to ion $i+1$, $\Gamma_i$ is the photoionization rate (s$^{-1}$) from ion $i$ to ion $i+1$, and $X_i$ is the CX reaction rate from ion $i$ to ion $i-1$. The following ion species are included in our CX calculation: \ion{C}{6}--\ion{C}{7}, \ion{N}{7}--\ion{N}{8}, \ion{O}{8}--\ion{O}{9}, and \ion{Ne}{10}--\ion{Ne}{11}, which cover the soft X-ray band most relevant for CGM and GC observations. The rates $\Gamma_i$ are evaluated in the optically thin limit for an isotropic metagalactic ionizing background, adopting the Haardt \& Madau spectrum at redshift zero and neglecting local radiation sources.

The CX recombination term in Eq.~\eqref{eq:nei} is calculated as the sum of neutral-hydrogen and neutral-helium donor channels:
\begin{equation} \label{eq:cx_rate}
\begin{aligned}
    X_i &= \sum_{d\in\{\mathrm{H},\,\mathrm{He}\}} R_i^d,\\
    R_i^d &= n_i n_d v_{\mathrm{rel},d}\sigma_{i,d},
\end{aligned}
\end{equation}
where $R_i^d$ (cm$^{-3}$\,s$^{-1}$) is the volumetric reaction rate for ion $i$ with neutral donor $d$, $X_i$ is the rate obtained by summing over H and He donors, $n_d$ is the donor number density, $\sigma_{i,d}$ is the velocity-dependent CX cross section, and $v_{\mathrm{rel},d}$ is the ion-donor relative velocity. The cross sections $\sigma_{i,d}(v)$ are taken from the ACX2 database \citep{foster2020pyatomdb}, which incorporates velocity-dependent theoretical charge-exchange cross sections from the Kronos database \citep{mullen2016charge}. The neutral hydrogen abundance is approximated assuming local ionization equilibrium, while neutral helium is followed explicitly in the NEI network. We discuss the limitations of the hydrogen approximation in \S\ref{subsec:caveats}.

\subsection{Ion-Neutral Relative Velocity} \label{subsec:rel_velocity}

Eq.~\eqref{eq:cx_rate} only contains one subgrid quantity requiring modeling: the effective relative velocity $v_{\mathrm{rel},d}$. Since a single-fluid calculation does not separately evolve ion and neutral velocities, we estimate the effective ion-donor relative velocity as the quadrature sum of thermal and unresolved turbulent motions:
\begin{equation}
    v_{\mathrm{rel},d} = \sqrt{v_{\mathrm{th},d}^2 + v_{\mathrm{turb},d}^2}.
\end{equation}

\emph{Thermal component:}
For a Maxwellian distribution, the mean thermal relative velocity between an ion of mass $m_i$ and a neutral donor of mass $m_d$ is
\begin{equation}
    v_{\mathrm{th},d} = \sqrt{\frac{8 k_\mathrm{B} T}{\pi \mu_{\mathrm{red},d}}},
\end{equation}
where $T$ is the local temperature and $\mu_{\mathrm{red},d} = m_i m_d / (m_i + m_d)$ is the reduced mass. For heavy ions, this approaches the donor thermal speed. The cell temperature is a single-fluid average, so potentially unresolved multi-temperature structure below the grid scale remains an approximation.

\emph{Subgrid turbulent component:}
We estimate unresolved turbulent motions using Kolmogorov scaling, $\delta v(r) \propto \varepsilon^{1/3} r^{1/3}$, where $\varepsilon$ is the turbulent energy dissipation rate per unit mass. This provides a controlled extrapolation below the grid scale and is broadly consistent with CGM velocity structure functions \citep{chen2023cosmic}. Applying this scaling requires two ingredients: a grid-scale velocity fluctuation to anchor the extrapolation, and the length scale of the CX interaction itself, to which the extrapolation is carried out. We describe each in turn below.

The resolved cell-scale velocity fluctuation, which anchors the extrapolation, is estimated from the local strain rate:
\begin{equation}
    \delta v(\Delta x) = S \Delta x,
\end{equation}
with
\begin{equation} \label{eq:strain_rate}
    S = \sqrt{2 S_{ij} S_{ij}}, \qquad S_{ij} = \frac{1}{2}\left(\frac{\partial v_i}{\partial x_j} + \frac{\partial v_j}{\partial x_i}\right),
\end{equation}
where $\Delta x$ is the cell size, $S_{ij}$ is the strain-rate tensor, and $v_i$ is the velocity component along the $i$-th axis. This is analogous to a filter-scale fluctuation in large-eddy simulation models \citep{pope2000turbulent,sagaut2006large}.

The target length scale for the extrapolation is set by the CX interaction itself -- the mean free path of ion $i$ against charge exchange with neutral donor $d$:
\begin{equation}
    \lambda_{\mathrm{CX},d} = \frac{1}{n_d \sigma_{i,d}},
\end{equation}
Here the dependence on ion species $i$ is implicit in $\lambda_{\mathrm{CX},d}$, which determines the distance an ion typically travels before reacting with that donor. In addition, because the cross section $\sigma_{i,d}$ depends on the relative velocity, and the velocity estimate in turn uses the mean free path, we solve $v_{\mathrm{rel},d}$ and $\lambda_{\mathrm{CX},d}$ iteratively, starting from the grid-scale estimate $\delta v(\Delta x)$. With $\lambda_{\mathrm{CX},d}$ determined, we compare it to the grid scale to decide how the Kolmogorov extrapolation should be applied: when $\lambda_{\mathrm{CX},d} < \Delta x$, we extrapolate the turbulent velocity from the grid scale down to the interaction scale via Kolmogorov scaling,
\begin{equation} \label{eq:v_turb}
    v_{\mathrm{turb},d} = \delta v(\Delta x) \left( \frac{\lambda_{\mathrm{CX},d}}{\Delta x} \right)^{1/3}.
\end{equation}
When $\lambda_{\mathrm{CX},d} \geq \Delta x$, the grid-scale velocity fluctuation is already resolved, so we set $v_{\mathrm{turb},d} = \delta v(\Delta x)$.

We finally note that the Kolmogorov scaling may be modified at small scales by viscosity, magnetic fields, etc. However, in the simulations analyzed here, the thermal component always dominates $v_{\mathrm{rel},d}$ in CX-emitting gas, while $v_{\mathrm{turb},d}$ is comparatively small, limiting the quantitative influence of this cascade extrapolation.

\subsection{Soft X-ray Emission from Charge Exchange} \label{subsec:emission}

CX reactions produce characteristic X-ray line emission as the captured electron cascades from its initial highly excited state to the ground configuration of the recombined ion. The dominant CX emission lines in the CGM context arise from the same H-like and He-like ions of C, N, O, and Ne that carry appreciable CX cross sections. Throughout this paper, soft X-ray refers to the 0.5--2~keV band, which contains the bulk of the observationally accessible CX and thermal line emission.

The total volumetric CX emissivity, summed over all emitted lines regardless of energy, is
\begin{equation}
    \Lambda_\mathrm{CX} = \sum_{i,\,d,\,l} R_i^d\,y_{il}^d\,E_{il}^d,
    \label{eq:cx_emissivity}
\end{equation}
where $R_i^d$ is given by Eq.~\eqref{eq:cx_rate}, $y_{il}^d$ is the line-yield fraction for transition $l$, and $E_{il}^d$ is the corresponding photon energy. The line yields $y_{il}^d$ are taken directly from ACX2 \citep{foster2020pyatomdb}. Restricting the sum in Eq.~\eqref{eq:cx_emissivity} to lines $l$ within the 0.5--2~keV soft X-ray band gives the band-limited CX emissivity,
\begin{equation}
    \epsilon_\mathrm{CX} = \sum_{i,\,d,\,l\,\in\,[0.5,2]\,\mathrm{keV}} R_i^d\,y_{il}^d\,E_{il}^d,
    \label{eq:cx_emissivity_band}
\end{equation}
a subset of the total $\Lambda_\mathrm{CX}$. This is the quantity used throughout the paper, together with the same band-limited thermal soft X-ray emissivity $\epsilon_\mathrm{th}$.

\section{Numerical Methods} \label{sec:methods}

We implement the CX model in the {\small FLASH} code \citep{fryxell2000flash}. The code solves inviscid ideal hydrodynamics with $\gamma=5/3$; the equations and numerical scheme are identical to those of \citet{ji2019simulations}. Ion fractions governed by Eq.~\eqref{eq:nei} are evolved in parallel with the hydrodynamics, using the {\small FLASH} NEI solver, with photoionization incorporated by \citet{ji2019simulations}. The evolution of the gas energy equation, following \citet{ji2019simulations}, uses pre-tabulated metallicity- and density-dependent equilibrium cooling tables generated with {\small Cloudy} \citep{ferland20172017} for computational efficiency, and is evolved with the exact integration cooling scheme of \citet{townsend2009exact}. CX radiation does not enter as a cooling term in the thermal energy equation, since the emitted photons draw primarily on the ionization potential energy carried by highly charged ions rather than directly on the local thermal energy.

The three-dimensional simulations are initialized with a single cold-hot interface and a shear velocity that drives the development of a radiative turbulent mixing layer through the KH instability, following the setups of \citet{ji2019simulations} and \citet{yang2023radiative}: the $x$-axis is parallel to the shear flow and $z$ is normal to the initial interface. Stationary cold gas at $z>0$ and hot gas moving at $v_\mathrm{hot}$ at $z<0$ begin in pressure equilibrium; hyperbolic-tangent density and velocity profiles join the phases across a layer of half-width $a$ (which is set to $2.5\%$ of the domain width). A transverse sinusoidal perturbation of amplitude $0.01 v_\mathrm{hot}$ seeds the KH instability. Outflowing and periodic boundary conditions are applied in the $z$ and $x$ directions, respectively. The standard grid is $128^2\times384$ cells and $256^2\times768$ is used for convergence tests.

Tab.~\ref{tab:run_parameters} lists the physical parameters of the simulations. We consider two representative environments: the CGM and the GC, characterized by different densities, temperatures and density contrasts between the cold and hot phases. Runs also vary metallicity and shear velocity. In particular, to isolate the response to radiative cooling with all other parameters fixed, we also run a few cooling-strength variants by multiplying the entire fiducial solar-metallicity cooling curve by a constant prefactor $f_\Lambda=\Lambda/\Lambda_\mathrm{Z=1}$, e.g., \texttt{CGM-Z1-Lambda0.1} and \texttt{CGM-Z1-Lambda10} represent runs with 0.1 and 10 times the fiducial solar-metallicity cooling curve, respectively.

\begin{table*}[ht!]
\caption{Simulation parameters. Number densities are hydrogen number densities; $v_\mathrm{hot}$ is the initial velocity of the hot lower layer relative to the stationary cold upper layer. Symbols \texttt{Z}, \texttt{v}, \texttt{HR}, and \texttt{Lambda} in the aliases denote variants in metallicity, hot-phase velocity, high resolution, and cooling strength prefactor, respectively; photoionization is turned off except in the \texttt{CGM-Z0.3-Ph} run.}
\label{tab:run_parameters}
\centering
\setlength{\tabcolsep}{3pt}
\begin{tabular}{llcccccccc}
\toprule
Alias & Environment & $n_\mathrm{cold}$ & $T_\mathrm{cold}$ & $n_\mathrm{hot}$ & $T_\mathrm{hot}$ & $v_\mathrm{hot}$ & $Z$ & Box size & Grid \\
 & & (cm$^{-3}$) & (K) & (cm$^{-3}$) & (K) & (km s$^{-1}$) & & (pc) & \\
\midrule
{\tt CGM-Z0.1} & CGM-like & $10^{-2}$ & $10^4$ & $10^{-4}$ & $10^6$ & $100$ & $0.1\,Z_\odot$ & $100^2\times300$ & $128^2\times384$ \\
{\tt CGM-Z0.3-v25} & CGM-like & $10^{-2}$ & $10^4$ & $10^{-4}$ & $10^6$ & $25$ & $0.3\,Z_\odot$ & $100^2\times300$ & $128^2\times384$ \\
{\tt CGM-Z0.3-v50} & CGM-like & $10^{-2}$ & $10^4$ & $10^{-4}$ & $10^6$ & $50$ & $0.3\,Z_\odot$ & $100^2\times300$ & $128^2\times384$ \\
{\tt CGM-Z0.3} & CGM-like & $10^{-2}$ & $10^4$ & $10^{-4}$ & $10^6$ & $100$ & $0.3\,Z_\odot$ & $100^2\times300$ & $128^2\times384$ \\
{\tt CGM-Z0.3-Ph} & CGM-like & $10^{-2}$ & $10^4$ & $10^{-4}$ & $10^6$ & $100$ & $0.3\,Z_\odot$ & $100^2\times300$ & $128^2\times384$ \\
{\tt CGM-Z0.3-HR} & CGM-like & $10^{-2}$ & $10^4$ & $10^{-4}$ & $10^6$ & $100$ & $0.3\,Z_\odot$ & $100^2\times300$ & $256^2\times768$ \\
{\tt CGM-Z0.3-v150} & CGM-like & $10^{-2}$ & $10^4$ & $10^{-4}$ & $10^6$ & $150$ & $0.3\,Z_\odot$ & $100^2\times300$ & $128^2\times384$ \\
{\tt CGM-Z1} & CGM-like & $10^{-2}$ & $10^4$ & $10^{-4}$ & $10^6$ & $100$ & $Z_\odot$ & $100^2\times300$ & $128^2\times384$ \\
{\tt CGM-Z1-Lambda[$f_\Lambda$]} & CGM-like & $10^{-2}$ & $10^4$ & $10^{-4}$ & $10^6$ & $100$ & $Z_\odot$ & $100^2\times300$ & $128^2\times384$ \\
{\tt GC-Z1} & GC-like & $10^{1}$ & $10^3$ & $10^{-2}$ & $10^6$ & $100$ & $Z_\odot$ & $1^2\times3$ & $128^2\times384$ \\
{\tt GC-Z1-HR} & GC-like & $10^{1}$ & $10^3$ & $10^{-2}$ & $10^6$ & $100$ & $Z_\odot$ & $1^2\times3$ & $256^2\times768$ \\
\bottomrule
\end{tabular}
\end{table*}

\section{Results} \label{sec:results}

\subsection{Spatial Morphology of the CX-Emitting Layer} \label{subsec:results_morphology}

\begin{figure*}[t]
    \centering
    \includegraphics[width=0.498\textwidth]{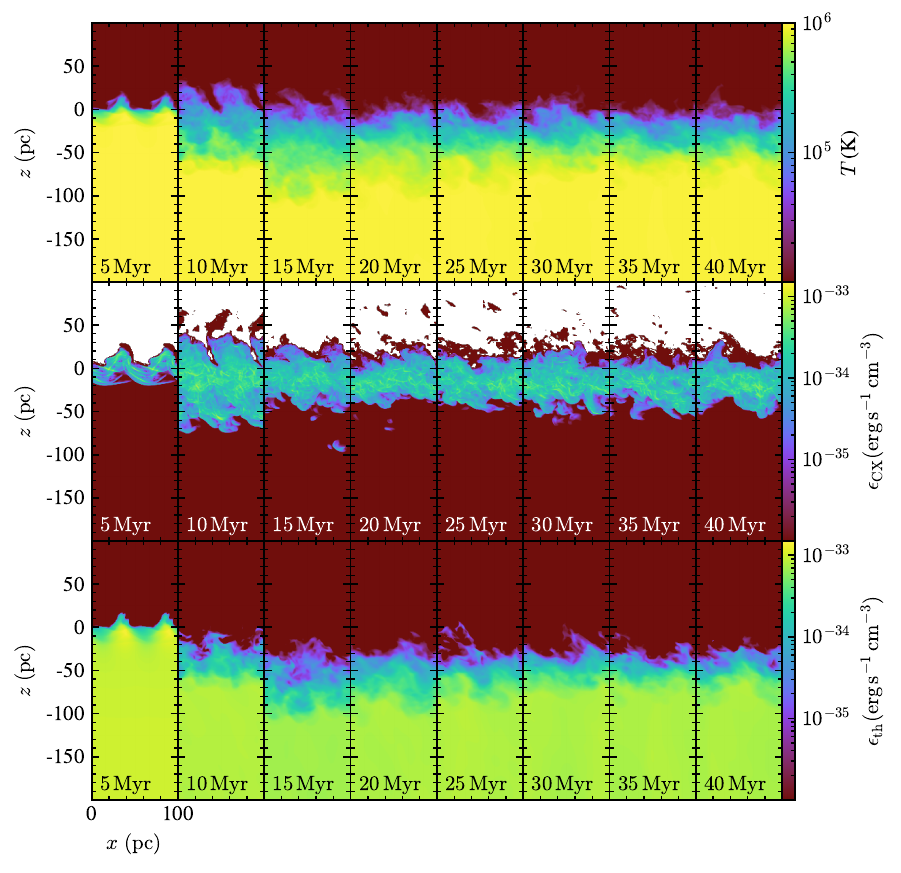}
    \includegraphics[width=0.498\textwidth]{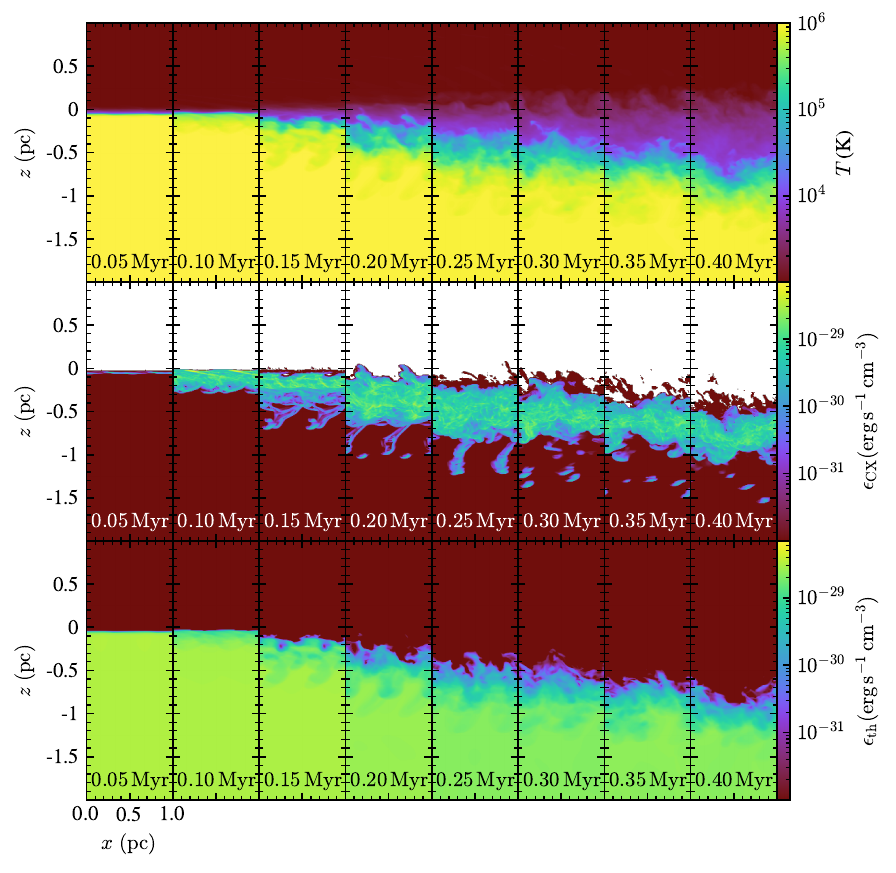}
    \caption{Time evolution in the {\tt CGM-Z0.3} (left) and {\tt GC-Z1} (right) runs, showing volume-weighted projections of temperature $T$ (top) and soft X-ray emissivities from CX $\epsilon_\mathrm{CX}$ (middle) and thermal emission $\epsilon_\mathrm{th}$ (bottom). CX emission is confined to the corrugated cold-hot mixing interface, while thermal soft X-ray emission traces the hot phase.}
    \label{fig:temp_cx_thermal}
\end{figure*}

We first compare {\tt CGM-Z0.3} and {\tt GC-Z1} as representative CGM and GC cases. Since {\tt GC-Z1} is scaled down by a factor of 100 in both length and time, we compare all timescales relative to the crossing time. Its larger density contrast ($10^3$ versus $10^2$) increases $t_\mathrm{KH}/t_\mathrm{cross}$ by $\sim\sqrt{10}$. Meanwhile, adopting $T_\mathrm{mix}\sim\sqrt{T_\mathrm{cold}T_\mathrm{hot}}$ and $n_\mathrm{mix}\sim\sqrt{n_\mathrm{cold}n_\mathrm{hot}}$, the mixing layer in {\tt GC-Z1} is both cooler and denser than in {\tt CGM-Z0.3}, and its metallicity is also higher ($Z_\odot$ versus $0.3Z_\odot$). Together, these reduce the absolute cooling time by a factor of a few hundred, or equivalently, reduce $t_\mathrm{cool}/t_\mathrm{cross}$ by a factor of a few. Cooling is therefore more competitive with KH growth in {\tt GC-Z1}. Consequently, {\tt CGM-Z0.3} develops more prominent KH structures at early times, whereas cooling has a stronger influence on the evolution of {\tt GC-Z1}.
 
The top row of Fig.~\ref{fig:temp_cx_thermal}, showing the time evolution of temperature in {\tt CGM-Z0.3} and {\tt GC-Z1}, confirms the expectations above. In both runs, the KH instability develops and drives turbulence at the cold-hot interface, producing a corrugated mixing layer at intermediate temperatures that reaches a quasi-steady state after a few KH growth times. The temperature maps reveal more pronounced KH roll-up in {\tt CGM-Z0.3} during the early evolution, consistent with its longer cooling time relative to the KH growth time. In {\tt GC-Z1}, coherent billows are less apparent, as expected when cooling acts more rapidly relative to KH growth, and the mixing layer progressively shifts downward into the initially hot lower half of the domain. This downward displacement is consistent with cold-gas growth via cooling and condensation of entrained hot gas \citep{ji2019simulations}.

The second and third rows of Fig.~\ref{fig:temp_cx_thermal} show the corresponding CX and thermal soft X-ray emissivities on the same color scale, enabling a direct comparison of the two emission mechanisms. Two key features stand out. First, the two mechanisms differ significantly in spatial morphology: CX emission closely tracks the development of the mixing layer, remaining confined to the cold-hot transition where cold or rapidly cooling gas contacts hot gas, whereas thermal soft X-ray emission instead traces the hot volume, filling the hot side of the domain and fading toward the mixing layer as the hot gas is diluted and cooled. Second, regarding the magnitudes, the CX emissivity is surprisingly strong, with the projected CX emission from the mixing layer comparable to, and in some regions even exceeding, the thermal emission from the hot phase. When the comparison is restricted to the mixing layer itself, CX emission is often brighter than thermal emission. Therefore, CX is potentially a significant, and even dominant, contributor to the soft X-ray emission from cold-hot interfaces.

\subsection{CX and Thermal Emissivity Distributions} \label{subsec:results_emissivity_distribution}

\begin{figure*}[t]
    \centering
    \includegraphics[width=0.98\textwidth]{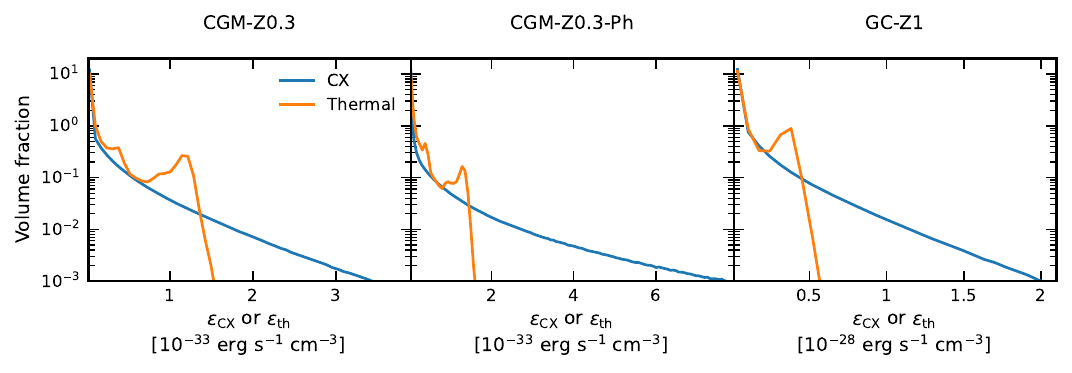}
    \caption{Volume-weighted distributions of the $0.5$--$2$~keV CX (blue) and thermal (orange) emissivities within the saturated turbulent mixing layers for {\tt CGM-Z0.3} (left), {\tt CGM-Z0.3-Ph} (middle), and {\tt GC-Z1} (right). To isolate the mixing layer from the imposed hot-phase flow, we exclude cells with $v_x>\langle v_x\rangle_T$, where $\langle v_x\rangle_T$ is the temperature-weighted mean shear velocity. Distributions are time-averaged over the saturated phase of each run ($80$--$100$~Myr for the CGM runs; $0.40$--$0.60$~Myr for {\tt GC-Z1}). CX extends to higher emissivities than thermal emission in all three runs.}
    \label{fig:emissivity_distribution}
\end{figure*}

Fig.~\ref{fig:emissivity_distribution} quantifies the volume-weighted distributions of CX and thermal soft X-ray emissivity in the mixing layers. We exclude cells with $v_x>\langle v_x\rangle_T$ to remove the imposed hot-phase flow and average over the saturated phase. The thermal distribution has a shoulder near its high-emissivity end, which is set by the imposed hot-phase flow, while the lower-emissivity gas includes material that has mixed and cooled out of the thermal soft X-ray-emitting range.
 
In all three runs, the CX distribution extends beyond the thermal cutoff, showing that a small volume of interface gas can have a higher soft X-ray emissivity than the hot phase. The comparison between \texttt{CGM-Z0.3} and \texttt{CGM-Z0.3-Ph} is especially useful because their initial gas parameters are identical. With photoionization, the CX tail extends to larger emissivities, while the thermal cutoff remains on a similar scale. This reinforces the picture of CX as a highly localized and bright diagnostic of the cold-hot interface.

\subsection{NEI and the Temperature Origin of CX Emission} \label{subsec:results_temperature_distribution}

\begin{figure*}[t]
    \centering
    \includegraphics[width=0.98\textwidth]{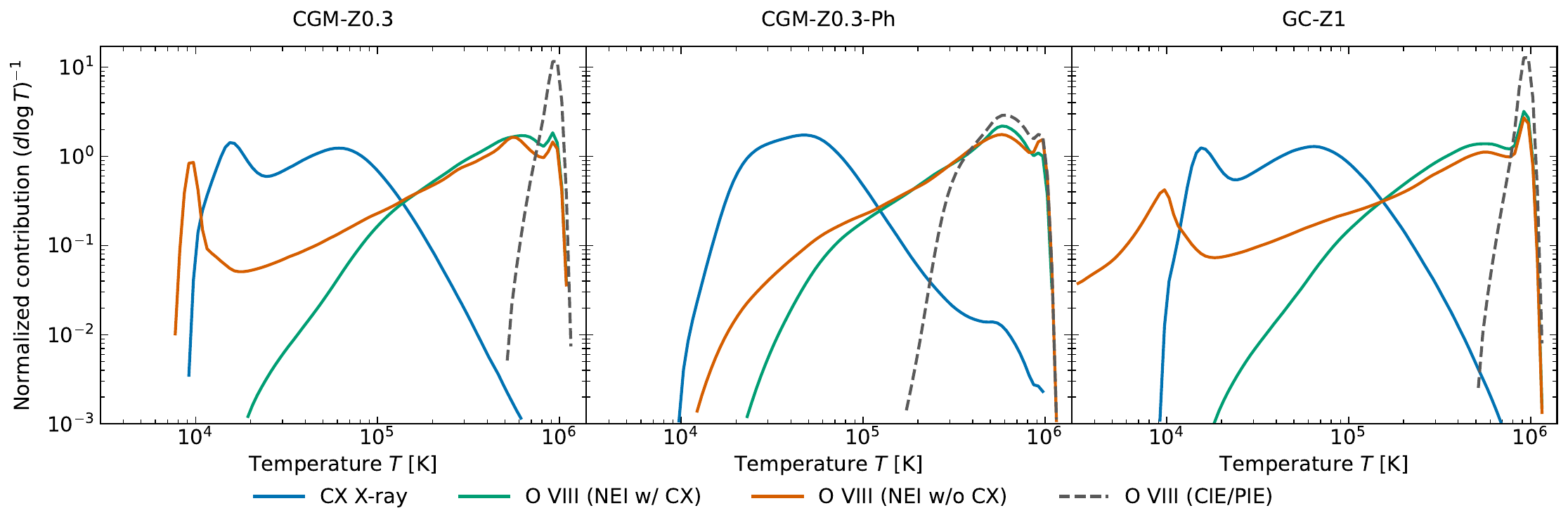}
    \caption{Normalized contributions per logarithmic temperature interval from the CX soft X-ray emissivity (solid blue) and \ion{O}{8} number density in the saturated mixing layers of {\tt CGM-Z0.3} (left), {\tt CGM-Z0.3-Ph} (middle), and {\tt GC-Z1} (right). Solid green and orange curves show \ion{O}{8} with and without CX, respectively. Dashed gray curves give the equilibrium reference: CIE in the left and right panels, and PIE in the middle panel. Each curve is normalized to its own integral; spatial selection and time averaging follow Fig.~\ref{fig:emissivity_distribution}. CX emission is concentrated at $T\sim10^4$--$10^5$~K, where the NEI distributions retain \ion{O}{8} far below the temperatures favored by the equilibrium reference. CX reduces the relative contribution from the cool gas.}
    \label{fig:temperature_distribution}
\end{figure*}

CX requires high-charge ions and neutral donors to coexist locally. We choose \ion{O}{8} as a representative high-charge ion that produces soft X-rays through CX, and examine its temperature distribution alongside that of the CX emissivity in Fig.~\ref{fig:temperature_distribution}. The equilibrium references are obtained by applying CIE or PIE ion fractions to the mixing-layer gas in post-processing, with PIE used for \texttt{CGM-Z0.3-Ph}. Each distribution is normalized to its own integral.

The equilibrium distributions (dashed gray) confine \ion{O}{8} mainly to high temperatures. In the runs without photoionization, CIE places it near $T\sim10^6$~K, at the hot end of the simulated temperature range. PIE extends the distribution moderately toward lower temperatures, but its contribution remains concentrated above $10^5$~K. Equilibrium ionization therefore predicts little \ion{O}{8} in the cool gas where neutral donors are abundant.

By contrast, with NEI included, the \ion{O}{8} distributions develop extended low-temperature tails. First consider the runs without CX (solid orange). Turbulent mixing and rapid cooling bring gas to low temperatures before electron recombination is complete. In the runs without photoionization, \ion{O}{8} can consequently accumulate in the cold phase near $10^4$~K, where the small free-electron abundance further slows recombination. We note that Fig.~\ref{fig:temperature_distribution} sums \ion{O}{8} number density over the gas volume in each temperature bin, so the temperature distribution curve depends on both the ion fraction and the mass of gas at that temperature; even a small ion fraction can therefore contribute significantly in the dense cold phase. In \texttt{CGM-Z0.3-Ph}, photoionization partially ionizes the cold gas, supplying free electrons that promote recombination of high-charge ions and suppress their accumulation at the cold end.

Enabling CX (solid green) further reduces the cold \ion{O}{8} contribution, as ions reaching neutral-rich gas capture electrons from neutral donors. The local CX and electron-recombination times for ion $i$ are $t_\mathrm{CX,i}=n_i/X_i=[\sum_d n_d v_{\mathrm{rel},d}\sigma_{i,d}]^{-1}$ and $t_\mathrm{rec,i}=(n_e\alpha_i)^{-1}$. For \ion{O}{8} in the brightest ($\gtrsim10\%\epsilon_\mathrm{CX,max}$) CX-emitting gas, $t_\mathrm{CX}$ is $\sim2\times10^5$~yr in {\tt CGM-Z0.3} and $\sim2\times10^3$~yr in {\tt GC-Z1}, compared with $t_\mathrm{rec}\sim2$--$3$~Myr and $\sim3\times10^3$--$5\times10^3$~yr, respectively. These estimates show that CX can remove high-charge ions faster than electron recombination in the emitting gas.

This surviving population sets the temperature origin of the CX emission (solid blue), which is concentrated at $T\sim10^4$--$10^5$~K. High-charge ions are abundant in the hot phase, but neutral donors are scarce there; in the cool interface, NEI allows these ions to persist and encounter neutrals. CX therefore traces the overlap between the transported ions and neutral gas, rather than the temperature range favored by equilibrium ionization.

\begin{figure}[t]
    \centering
    \includegraphics[width=\columnwidth]{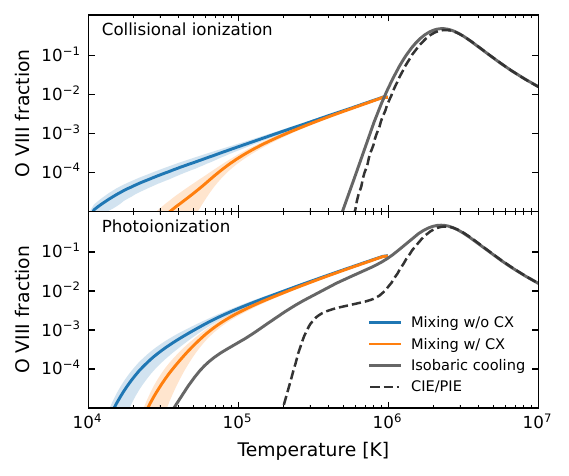}
    \caption{\ion{O}{8} fraction $f_\mathrm{OVIII}=n_\mathrm{OVIII}/n_\mathrm{O}$ versus temperature, without (top) and with (bottom) photoionization. Solid blue and orange curves show the mean fractions within temperature bins in the CGM mixing layers without and with CX, respectively; shaded bands span the 16th--84th percentiles. Solid gray curves show the time-dependent ion fraction in a uniform parcel cooling at constant external pressure, initialized at $T=10^7$~K and $n_\mathrm{H}=10^{-5}~\mathrm{cm^{-3}}$, at the same pressure as the CGM runs. Dashed dark gray curves show the corresponding CIE (top) and PIE (bottom) reference fractions. Mixing sustains more \ion{O}{8} in cool gas than isobaric radiative cooling alone, especially without photoionization; CX lowers the fraction at the cool end of the mixing layer.}
    \label{fig:oviii_isobaric}
\end{figure}

The temperature distribution in Fig.~\ref{fig:temperature_distribution} combines the ionization state with the density and volume distribution of the gas. To isolate the ionization state, Fig.~\ref{fig:oviii_isobaric} instead shows the mean of the ion fraction $f_\mathrm{OVIII}=n_\mathrm{OVIII}/n_\mathrm{O}$ within each temperature bin, where $n_\mathrm{O}$ is the total oxygen nuclei density. We also compare the results with a gas parcel evolving along a prescribed cooling trajectory \citep{oppenheimer2013nonequilibirum,gnat2017time,kumar2025nonequilibrium}. We adopt an isobaric trajectory because the subsonic mixing layers are approximately pressure balanced \citep{ji2019simulations}. Starting from $T=10^7$~K and $n_\mathrm{H}=10^{-5}~\mathrm{cm^{-3}}$, at the same pressure as the CGM runs, we evolve the parcel's specific enthalpy through $dh/dt=-\mathcal{L}/\rho$, where $\mathcal{L}$ is the net volumetric energy-loss rate and $\rho$ is the mass density, accounting for compression work and increasing the density as the gas cools to maintain constant external pressure.

Without photoionization (top panel), the isobarically cooling parcel remains close to CIE at high temperatures and develops only a limited low-temperature extension. The mean mixing-layer ion fractions, by contrast, extend throughout $T\sim10^4$--$10^5$~K, far below the temperature range in which the cooling parcel retains significant \ion{O}{8}. The shaded bands show the 16th--84th percentile scatter within each temperature bin. With photoionization (bottom panel), both PIE and isobaric cooling sustain \ion{O}{8} at lower temperatures, and the cooling parcel develops a more extended NEI tail. The mixing-layer fractions nevertheless remain above the isobaric-cooling curve over the shared low-temperature range. In both panels, CX reduces the cool-gas fraction, but does not erase the excess associated with mixing. The comparison thus shows that a single radiative cooling trajectory does not reproduce the ionization state of a turbulent interface: turbulence alters the temperature distribution of ion fractions through mixing of the phases as well as through mixing-induced cooling, and continuously supplies ions from the hot phase. The turbulent mixing-modified ionization history allows high-charge ions to coexist with neutral donors even after CX losses are included.

\subsection{Dependence on Metallicity, Shear Velocity, and Resolution} \label{subsec:velocity_metallicity_resolution}

\begin{figure}[!t]
    \centering
    \includegraphics[width=\columnwidth]{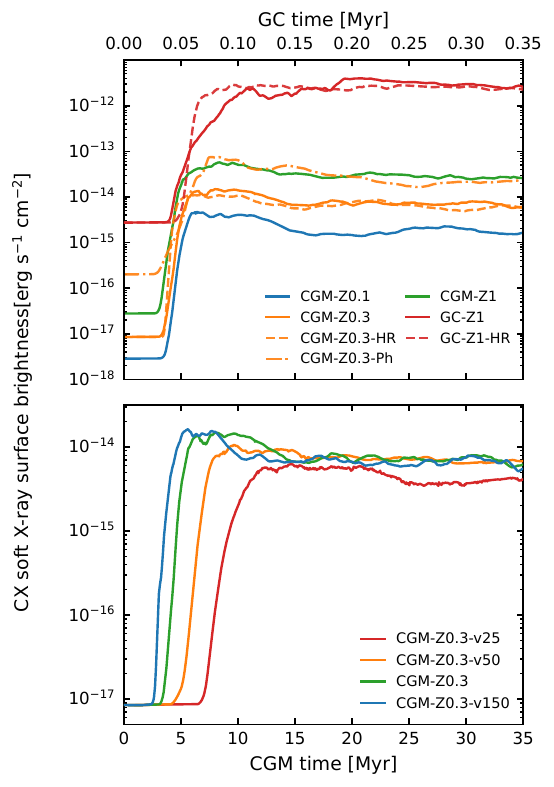}
    \caption{Time evolution of the CX soft X-ray surface brightness $Q_\mathrm{CX}$ defined by Eq.~\eqref{eq:cx_surface_brightness}. Top: {\tt CGM-Z0.1} (solid blue), {\tt CGM-Z0.3} (solid orange), {\tt CGM-Z1} (solid green), and {\tt GC-Z1} (solid red). Dashed orange and red curves show the corresponding high-resolution (HR) runs; the dash-dotted orange curve shows {\tt CGM-Z0.3-Ph}. CGM and GC times are read from the lower and upper axes, respectively. Bottom: the $0.3\,Z_\odot$ CGM shear sequence with $v_\mathrm{hot}=25$ (red), $50$ (orange), $100$ (green), and $150~\mathrm{km~s^{-1}}$ (blue).}
    \label{fig:Q_dependence}
\end{figure}

Since our turbulent mixing layers are initialized in a plane-parallel geometry, we define the area-averaged CX surface brightness as
\begin{equation}
Q_\mathrm{CX} = \int \left\langle \epsilon_\mathrm{CX}(l)\right\rangle_A dl,
\label{eq:cx_surface_brightness}
\end{equation}
where $l$ is the direction perpendicular to the velocity shear and $\langle\cdot\rangle_A$ denotes an average over the cross-sectional area perpendicular to $l$. Fig.~\ref{fig:Q_dependence} shows the time evolution of the CX surface brightness $Q_\mathrm{CX}$, where all of the runs follow the same qualitative evolution: the CX surface brightness first rises rapidly as the seeded KH instability corrugates the cold-hot interface and drives mixing, thereby increasing the initial contact between high-charge ions and neutral donors. It then approaches a saturation stage once mixing and cooling reach a dynamical balance, consistent with previous studies on radiative turbulent mixing layers \citep{ji2019simulations,tan2021radiative,yang2023radiative}.

The top panel shows a strong dependence of the saturated CX signal on metallicity and CGM/GC environment. Within the CGM sequence, increasing $Z$ both supplies more metal ions capable of producing soft X-ray CX lines and strengthens radiative cooling. The latter enhances cooling-induced entrainment and mixing, further increasing the overlap between high-charge ions and neutral gas; the cooling dependence is isolated and discussed in detail in \S\ref{subsec:cooling_strength}. At the same solar metallicity, {\tt GC-Z1} is brighter than {\tt CGM-Z1} by approximately two orders of magnitude. This is consistent with the higher hot-phase ion density in the GC setup, although its density contrast and physical scale also differ from those of the CGM runs.

The dash-dotted orange curve isolates the effect of photoionization by comparison with the solid orange curve. The saturated $Q_\mathrm{CX}$ in \texttt{CGM-Z0.3-Ph} is a factor of a few above that in \texttt{CGM-Z0.3}, consistent with the more extended high-emissivity tail in Fig.~\ref{fig:emissivity_distribution}. Although photoionization reduces the neutral donor abundance, it maintains higher metal charge states in the warm phase, consistent with the larger low-temperature \ion{O}{8} fractions in Fig.~\ref{fig:oviii_isobaric}. This suggests that as long as the neutral column remains sufficient to capture the incoming ions, the integrated CX emission is limited by ion supply rather than donor density, so with photoionization, a larger supply of high-charge ions can outweigh the reduction in donor density.

Regarding the numerical resolution dependence, the high-resolution runs (dashed curves) develop more rapidly during the initial KH growth stage, but converge closely to the standard-resolution runs after saturation. This behavior resembles earlier convergence studies, which found resolution-dependent transient KH growth but much weaker resolution dependence in saturated integral quantities \citep{ji2019simulations,tan2021radiative,yang2023radiative}. However, convergence of an integrated quantity does not establish convergence of the underlying phase structure; \citet{lancaster2026meaning} identify the turbulent Field length as an important resolution scale for the latter. Our comparison supports the robustness of saturated $Q_\mathrm{CX}$ over the tested factor of two in spatial resolution, while the measured interface area remains resolution dependent (\S\ref{subsec:3d_structure}).

The bottom panel of Fig.~\ref{fig:Q_dependence} shows that the imposed shear velocity primarily controls the onset of the mixing layer: faster shear produces faster KH growth and an earlier rise in $Q_\mathrm{CX}$. By contrast, the saturated surface brightness depends only weakly on $v_\mathrm{hot}$, apart from the modestly lower value in the slowest \texttt{CGM-Z0.3-v25} run. This is because, although changing the shear changes the turbulent velocity, the CX collision speed $v_\mathrm{rel}$ in these subsonic mixing layers is dominated by the thermal component and is thus insensitive to the turbulent velocity and hence to the shear velocity.

Based on the saturated $Q_\mathrm{CX}$ of the runs in Fig.~\ref{fig:Q_dependence}, we provide an approximate scaling relation for the saturated CX surface brightness from a single radiative turbulent mixing layer in the weak-cooling regime:
\begin{equation}
    Q_\mathrm{CX}\simeq
    Q_0
    \left(\frac{Z}{Z_\odot}\right)^{1.1}
    \left(\frac{n_\mathrm{H,hot}}{10^{-2}\ \mathrm{cm^{-3}}}\right)
    \mathrm{erg\,s^{-1}\,cm^{-2}},
\label{eq:Qcx_scaling}
\end{equation}
where $Q_0\sim4\times10^{-12}$. The single photoionization run demonstrates an enhancement factor of $\sim 5$ at $Z=0.3Z_\odot$. The scaling is calibrated for $T_\mathrm{hot}=10^6\,\mathrm{K}$, the density contrasts and physical scales covered by our simulations, and $v_\mathrm{hot}\sim10$--$10^2\,\mathrm{km\,s^{-1}}$. The fit includes the GC case with its higher gas density, and we find a proportional scaling on $n_\mathrm{H,hot}$, but only a weak dependence on $n_d$, so we omit an explicit $n_d$ dependence from Eq.~\eqref{eq:Qcx_scaling}. This weak dependence on $n_d$ can be understood from the short CX mean free path $\lambda_{{\rm CX},d}$. For the CX reaction layer whose thickness is set by \(\lambda_{{\rm CX},d}\), integration of the local emissivity across the layer gives
\begin{equation}
Q_{{\rm CX},i}
\sim n_i n_{d}\sigma_i v_{\rm rel}\overline{E}_i
\lambda_{{\rm CX},d}
\sim n_i v_{\rm rel}\overline{E}_i,
\end{equation}
where $\overline{E}_i$ is the mean energy emitted per CX reaction of ion $i$ including all yielded lines in the soft X-ray band. Thus, in the optically thick CX limit, the increase of neutral density raises the volume emissivity but reduces the reaction-layer thickness by the same factor. Once the neutral column is sufficient to consume the incident high-charge ions, adding neutral donors no longer increases the number of CX reactions per unit area, but instead confines them to a thinner layer. The surface brightness is therefore ion-supply limited rather than donor-density limited, consistent with the photoionization result discussed above.

\subsection{Cooling-strength Dependence of CX and Thermal Surface Brightness} \label{subsec:cooling_strength}

\begin{figure}[t]
    \centering
    \includegraphics[width=\columnwidth]{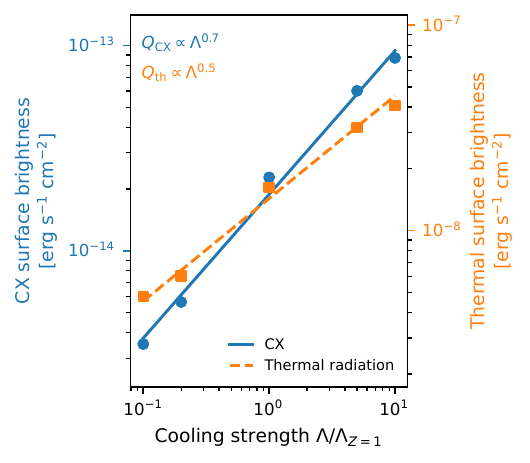}
    \caption{Saturated CX soft X-ray surface brightness (blue circles; left $y$-axis) and bolometric thermal cooling surface brightness (orange squares; right $y$-axis) as functions of the normalized cooling coefficient, for the {\tt CGM-Z1} cooling-strength sequence. Solid and dashed lines show power-law fits, $Q_\mathrm{CX}\propto \Lambda^{0.7}$ and $Q_\mathrm{th}\propto \Lambda^{0.5}$, respectively.}
    \label{fig:cooling_strength_scaling}
\end{figure}

Fig.~\ref{fig:cooling_strength_scaling} shows how the saturated CX soft X-ray surface brightness $Q_\mathrm{CX}$ and the total thermal cooling surface brightness $Q_\mathrm{th}$ depend on cooling strength, where $Q_\mathrm{th}$ is defined as
\begin{equation}
    Q_\mathrm{th}
    \equiv \int \left\langle n_e n_\mathrm{H}\Lambda(T)\right\rangle_A dl ,
\label{eq:thermal_cooling_surface_brightness}
\end{equation}
i.e., the bolometric radiative loss per projected area. The measured scaling $Q_\mathrm{th}\propto \Lambda^{1/2}$ arises from the entrainment scaling $v_\mathrm{in}\propto\Lambda^{1/2}$ found in the same weak-cooling regime \citep{ji2019simulations,tan2021radiative}, where $v_\mathrm{in}$ is the hot-phase entrainment velocity. In a saturated mixing layer, the mixed gas radiates approximately the enthalpy flux advected into the layer by entrainment, $Q_\mathrm{th}\sim P_\mathrm{hot} v_\mathrm{in}$, where $P_\mathrm{hot}$ is the hot-phase pressure.

The CX surface brightness rises somewhat more steeply, approximately $Q_\mathrm{CX}\propto \Lambda^{0.7}$. To identify the origin of this excess, we decompose Eq.~\eqref{eq:cx_emissivity_band} into its yield-weighted channel contributions, $R_i^d\sum_l y_{il}^d E_{il}^d$, with the line sum restricted to $0.5$--$2$~keV and $R_i^d=n_i n_d v_{\mathrm{rel},d}\sigma_{i,d}$ from Eq.~\eqref{eq:cx_rate}, which separates the ion-neutral density overlap from the collision factor $v_{\mathrm{rel},d}\sigma_{i,d}$. Summing over all included CX channels shows that the variation with cooling strength is dominated by the ion-neutral overlap $n_i n_d$, while changes in $v_\mathrm{rel}$ and $\sigma$ contribute only marginally.\footnote{Although the entrainment velocity $v_\mathrm{in}$ increases with cooling strength ($v_\mathrm{in} \propto \Lambda^{1/2}$), which increases the turbulent velocity, $v_\mathrm{rel}$ is already dominated by the thermal component in a subsonic turbulent mixing layer and is thus insensitive to the turbulent component (see \S\ref{subsec:velocity_metallicity_resolution}). Also, the cross section $\sigma$ generally decreases with increasing $v_\mathrm{rel}$, further suppressing the contribution of the product $v_\mathrm{rel}\sigma$ to the overall scaling.}

The dominance of the ion-neutral overlap ties this scaling to the NEI transport discussed in \S\ref{subsec:results_temperature_distribution}: CX consumes high-charge ions wherever they reach neutral-rich gas. The CX luminosity in both the CGM and GC therefore depends on the supply of these ions and on the true area of cold-hot contact over which they react. Turbulent entrainment and interface corrugation together set the integrated ion-neutral overlap $n_i n_d$, consistent with its dominant contribution to the cooling-strength dependence and the comparatively weak role of $v_\mathrm{rel}\sigma$. The entrainment velocity $v_\mathrm{in}$ sets both the incoming enthalpy flux underlying $Q_\mathrm{th}\propto\Lambda^{1/2}$ and the supply of high-charge ions, while the modestly steeper $Q_\mathrm{CX}\propto\Lambda^{0.7}$ scaling is consistent with additional changes in ion survival. Stronger cooling can bring ions into neutral-rich gas before electron recombination is complete, enhancing the overlap beyond that expected from the entrainment flux alone.

\subsection{3-dimensional Structure of the CX-Emitting Layer} \label{subsec:3d_structure}

\begin{figure*}[t]
    \centering
    \begin{interactive}{js}{sxcx_temperature_surface.zip}
    \includegraphics[width=\textwidth]{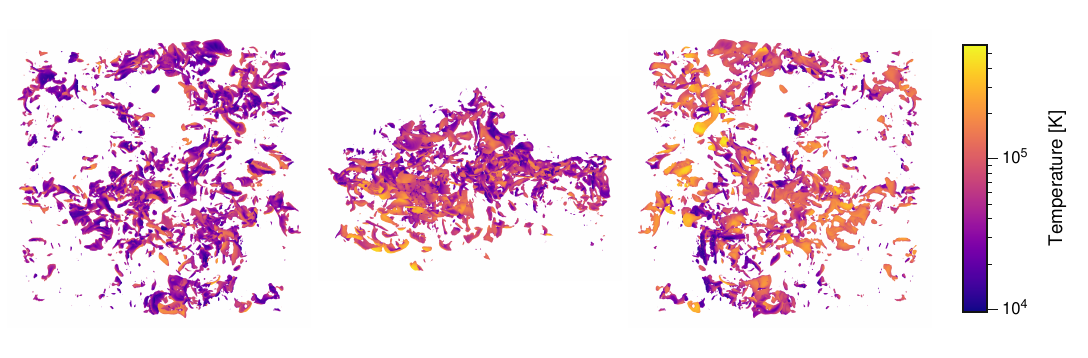}
    \end{interactive}
    \caption{Three orthographic views of the isosurface $\epsilon_\mathrm{CX}=10^{-33}~\mathrm{erg~s^{-1}~cm^{-3}}$ in the {\tt CGM-Z0.3-HR} run at $t=33\,\mathrm{Myr}$ within the saturated stage, colored by the local gas temperature. The left panel is a top-down view normal to the initial interface, seen from the cold phase; the middle panel is an edge-on view parallel to the interface; and the right panel is a bottom-up view normal to the interface, seen from the hot phase. The CX-emitting region forms fragmented, highly corrugated sheets and filaments that occupy a geometrically thin mixing layer while providing a large cold-hot contact area. An interactive 3D version of this figure is available at \url{https://skfb.ly/pO8Yw}.}
    \label{fig:3d_structure}
\end{figure*}

Fig.~\ref{fig:3d_structure} reveals the three-dimensional geometry underlying the projected CX maps of Fig.~\ref{fig:temp_cx_thermal}. Rather than forming a single smooth boundary between the cold and hot phases, the CX emissivity isosurface breaks up into a complex network of ribbon-like sheets, filaments, and disconnected patches. In the top-down and bottom-up views, these structures cover a substantial fraction of the projected interface but leave large gaps, showing that emission at $\epsilon_\mathrm{CX}=10^{-33}~\mathrm{erg~s^{-1}~cm^{-3}}$ is strongly intermittent within the interface plane. The edge-on view, by contrast, shows that these structures remain confined to a relatively narrow band while folding and overlapping within it. The CX-emitting ``layer'' is therefore thin in its overall extent, yet its true surface area is much larger than that of a smooth planar interface with the same projected footprint.

To quantify this corrugation, we measure the area $A_\mathrm{CX}$ of the CX-emissivity isosurface and define the corrugation factor for CX-emitting layers as
\begin{equation}
    \mathcal{C}_\mathrm{CX}
    \equiv \frac{A_\mathrm{CX}}{A_\mathrm{proj}},
    \label{eq:corrugation_factor}
\end{equation}
where $A_\mathrm{proj}$ is the projected area of the initial plane-parallel interface, and $A_\mathrm{CX}$ is the half-area of the CX isosurface at 1\% of the maximum CX emissivity in each run, measured using the \texttt{marching\_cubes} algorithm \citep{virtanen2020scipy}. The values of $\mathcal{C}_\mathrm{CX}$ averaged over the saturated stage are $8.40$ ($10.93$) for {\tt CGM-Z0.3} ({\tt CGM-Z0.3-HR}) and $6.92$ ($8.64$) for {\tt GC-Z1} ({\tt GC-Z1-HR}). The larger corrugation factors at higher resolution indicate that progressively smaller folds are being resolved, as expected for a fractal interface \citep{fielding2020multiphase}. Taking the grid cell scale $\ell$ as the effective smoothing scale, $A(\ell)\propto\ell^{2-D}$ and the factor-of-two increase from fiducial to high-resolution (\texttt{-HR}) simulations imply CX fractal dimensions of $D_\mathrm{CX}\simeq2.38$ and $2.32$ for the CGM and GC runs, respectively. These values are somewhat lower than the $D=2.5$ measured by \citet{fielding2020multiphase} for the cooling surface in turbulent mixing layers, but are consistent with the experimental range $D=2.1$--$2.4$, with a preferred value of $D=2.35$, for turbulent combustion fronts summarized by \citet{tan2021radiative}. Interestingly, both \citet{fielding2020multiphase} and \citet{tan2021radiative} focus on the strong-cooling regime, while our \texttt{CGM-Z0.3} run is clearly in the weak-cooling regime. We suspect that the CX surface can remain thin in a weak-cooling layer because CX constitutes a fast local reaction once high-charge ions encounter neutral gas, determined by $t_\mathrm{CX}$ in \S\ref{subsec:results_temperature_distribution}. This suggests that a short CX time can selectively illuminate a corrugated contact surface even when the radiative cooling region is broad.

The temperature coloring in Fig.~\ref{fig:3d_structure} further demonstrates that this surface is multiphase: in the top-down view from the cold side (left panel), the CX surface is weighted toward cooler gas (purple), whereas in the bottom-up view from the hot side (right panel), it is weighted toward warmer gas (yellow). Both views also contain overturned patches spanning the opposite end of the color scale, showing that turbulent motions fold the interface back on itself. Thus, CX traces a corrugated cold-hot contact surface with substantial turbulent overturning, rather than a smooth or volume-filling layer.

 \section{Discussion and Conclusions}
\label{sec:discussion_conclusions}

\subsection{Interface Geometry and the CX Contribution} \label{subsec:cx_dominance}

To relate the local interface emission to the luminosity of a multiphase system, we assume that the local CX reaction structure in simulated plane-parallel mixing layers is statistically valid even when the macroscopic distribution and geometry of the cold gas differ. With the saturated surface brightness $Q_\mathrm{CX}$ as the effective CX luminosity per unit macroscopic interface area, the total CX luminosity is then
\begin{equation}
    L_\mathrm{CX}
    \simeq Q_\mathrm{CX} A_\mathrm{int},
    \label{eq:Lcx_geometry}
\end{equation}
where $A_\mathrm{int}$ is the macroscopic interface area set by the sizes and shapes of the cold structures, before accounting for the local turbulent corrugation captured in our plane-parallel mixing-layer simulations. This corrugation, quantified by $\mathcal{C}_\mathrm{CX}$ in \S\ref{subsec:3d_structure}, is already absorbed into the measured $Q_\mathrm{CX}$ from simulations (defined per unit projected area) and should not be counted again in $A_\mathrm{int}$.

As indicated by the emissivity distributions in
\S\ref{subsec:results_emissivity_distribution},
thermal soft X-ray emission predominantly comes from the hot phase.
Its luminosity is approximately
\begin{equation}
    L_\mathrm{th}
    \simeq
    \overline{\epsilon}_\mathrm{th} f_\mathrm{hot} V,
    \label{eq:Lth_geometry}
\end{equation}
where $\overline{\epsilon}_\mathrm{th}$ is the volume-averaged
$0.5$--$2$~keV emissivity of the hot phase, $f_\mathrm{hot}$ is
its volume filling fraction, and $V$ is the system volume.
The luminosity ratio therefore becomes
\begin{equation}
    \frac{L_\mathrm{CX}}{L_\mathrm{th}}
    \simeq
    \frac{Q_\mathrm{CX}}{\overline{\epsilon}_\mathrm{th}}
    \frac{1}{f_\mathrm{hot}(V/A_\mathrm{int})},
    \label{eq:cx_geometry_general}
\end{equation}
where $V/A_\mathrm{int}$ has dimensions of length and represents the effective volume-to-interface scale of the multiphase medium: its inverse, $A_\mathrm{int}/V$, measures the cold-hot contact area available per unit system volume. It is a measure of the overall interface abundance: for example, a population of spherical cold clouds of equal radius $r_\mathrm{c}$, fully surrounded by hot gas and occupying a volume fraction $f_\mathrm{cold}$, has $V/A_\mathrm{int}=r_\mathrm{c}/(3f_\mathrm{cold})$. At a fixed cold-gas volume fraction, breaking the cold gas into smaller clouds increases their total contact area and decreases $V/A_\mathrm{int}$. We note that the local turbulent corrugation already included in $Q_\mathrm{CX}$ must still be excluded from $A_\mathrm{int}$.

Therefore, with a calibrated $Q_\mathrm{CX}$ and independent constraints on the hot phase, Eq.~\eqref{eq:cx_geometry_general} converts a spectroscopically measured CX-to-thermal luminosity ratio into a constraint on $V/A_\mathrm{int}$. At a fixed cold-gas volume fraction, smaller clouds or more complex shapes provide more contact area and a larger CX-to-thermal ratio. We caution that this estimate is based on plane-parallel layers where all CX comes from turbulent mixing, and may vary with the large-scale gas kinematics and cold-gas geometry. For instance, head-on collisions between cold clouds and hot winds may lead to strong CX production on the collision front without causing turbulent mixing. Future work will test this estimate in a broader range of multiphase systems, including cloud-crushing flows \citep{gronke2018growth,li2019survival,sparre2020interaction,abruzzo2022simple}, turbulent multiphase media \citep{gronke2021survival}, and galactic winds \citep{brueggen2016launching,schneider2017hydrodynamical}.

\subsection{Observational Implications} \label{subsec:observational_implications}

In the CGM, CX could trace the contact between cool clouds or filaments and surrounding hot gas. A locally bright interface may make only a small contribution to the integrated emission if its area is small relative to the hot volume. Conversely, a fragmented cold phase provides more contact area and can increase the relative CX signal. High-velocity clouds (HVCs) provide a possible test at localized CGM interfaces. UV absorption observations show cool HVC components accompanied by intermediate and high ions, consistent with interaction with surrounding hot gas \citep{tripp2022high}. Their boundaries and stripped fragments are therefore promising locations for spatially resolved CX searches. Enhanced forbidden emission in He-like triplets and high-$n$ cascades, together with spatial coincidence with neutral gas, can help distinguish CX from thermal emission. Future high-spectral-resolution X-ray missions such as HUBS \citep{cui2020hubs,bregman2023scientific} could use these diagnostics to constrain exchange between the cool and hot phases.

Galactic winds offer another promising setting: entrained clouds and filaments concentrate cold-hot contact area within the outflow, and observations of their boundaries can reduce dilution by unrelated hot gas along the sightline. Continuous interaction with the wind also supplies high-charge ions to these interfaces. XMM--Newton spectra of M82 and other nearby star-forming galaxies have shown line ratios consistent with a substantial CX contribution, although the inferred fraction remains model dependent and can be degenerate with a multi-temperature thermal plasma \citep{liu2011m82,zhang2014m82,wu2020outflows,fukushima2024revisiting}. Meanwhile, HST and JWST imaging resolves H$\alpha$-emitting clouds, PAH-bearing filaments, and dust embedded in the M82 superwind \citep{lopez2025observational,lopez2026jwst,cronin2026jwst}. Our calculations predict that the strongest CX emission should follow the surfaces of these cool structures rather than the volume-filling hot wind. Because the CX luminosity scales with interface area, such measurements would also constrain the unresolved fragmentation that controls cloud survival, mass loading, and entrainment \citep{gronke2020how,fielding2022structure,thompson2024theory}.

Finally, the GC contains dense molecular structures, warm ionized gas, supernova-heated plasma, and nuclear outflows in close proximity \citep{ponti2015xmm,kruijssen2014controls,heywood2019inflation}. Foreground absorption and the coexistence of thermal plasma, nonthermal emission, and X-ray reflection make the GC a challenging case, but they also provide strong morphological tests: CX-sensitive lines should trace cloud edges and folded contact surfaces rather than cloud interiors or the full hot volume. Detecting and modeling this component would constrain mass loading and entrainment in the nuclear outflow, as well as the exchange of material between cold clouds and the hot phase that ultimately regulates gas delivery toward Sgr~A$^*$.

\subsection{Conclusions} \label{subsec:conclusions}

Our main conclusions are as follows:
\begin{enumerate}[label=(\roman*)]
    \item \emph{Soft X-rays can originate in cool interface gas:} CX soft X-rays arise predominantly at $T\sim10^4$--$10^5$~K, where high-charge ions supplied by the hot phase encounter neutral donors. This effect is enhanced by photoionization under CGM-like conditions. CX illuminates the cold-hot interface, whereas thermal soft X-ray emission fills the hot volume. Within the mixing layer, CX can locally outshine thermal soft X-ray emission.

    \item \emph{Mixing-modified NEI is essential to CX:} Turbulent entrainment carries high-charge ions (e.g., \ion{O}{8}) from the hot phase into rapidly cooling gas faster than electron recombination can restore CIE or PIE. The high-charge ion fractions modified by turbulent mixing in cool/warm phases exceed not only their CIE or PIE reference, but also the fractions in a uniformly cooling isobaric parcel. Turbulent exchange and continued entrainment therefore matter to the ionization history. CX emission traces the continuing transport of high-charge ions across the cold-hot interface, rather than an equilibrium temperature phase.

    \item \emph{CX is ion-supply limited rather than donor-density limited at contact surfaces:} When the neutral column is sufficient to capture the incident ions, increasing the donor density raises the local emissivity but shortens the reaction layer, leaving no explicit donor-density dependence in its integrated brightness. This is further supported by the photoionization run: a larger supply of high-charge ions outweighs the reduction in neutral donors, leading to an overall increase in CX by a factor of a few.

    \item \emph{Strong dependence on cooling strength and weak dependence on velocity:} Stronger cooling, with a larger cooling coefficient $\Lambda$, increases the single-layer CX surface brightness as $Q_\mathrm{CX}\propto\Lambda^{0.7}$. This is slightly steeper than the thermal scaling $Q_\mathrm{th}\propto\Lambda^{0.5}$ expected from $v_\mathrm{in}\propto\Lambda^{1/2}$ in weak-cooling turbulent mixing layers, where $v_\mathrm{in}$ is the hot-phase entrainment velocity. The excess is driven mainly by enhanced ion-neutral overlap rather than by the collision factor $v_\mathrm{rel}\sigma$. Faster shear accelerates the onset of CX emission but only weakly changes its saturated surface brightness, because thermal motions dominate $v_\mathrm{rel}$ in these subsonic layers.

    \item \emph{CX interfaces are thin but highly corrugated:} CX emission forms intermittent sheets, filaments, and patches within a geometrically narrow mixing layer. The measured emitting-surface area exceeds the projected planar area by a factor $\mathcal{C}_\mathrm{CX}\simeq6.9$--$10.9$. Opposite views preferentially expose cooler and warmer faces, with overturned patches tracing turbulent folding. CX therefore directly maps the folded contact surface and provides a clean tracer of the cold-hot interface.

    \item \emph{The global CX contribution depends on multiphase geometry:} CX luminosity scales with the cold-hot interface area $A_\mathrm{int}$, whereas thermal soft X-ray luminosity scales with the hot-gas volume $V$. Their ratio therefore scales as $A_\mathrm{int}/V$ and increases as the effective volume-to-interface scale $V/A_\mathrm{int}$ decreases. Conversely, an observed CX-to-thermal luminosity ratio, combined with the CX surface brightness calibrated from simulations, determines $V/A_\mathrm{int}$. Given estimates of the system volume and cold-gas filling fraction, this constrains the available interface area and hence the degree of cold-gas fragmentation.

\end{enumerate}

\subsection{Caveats} \label{subsec:caveats}

Several limitations should be kept in mind when interpreting these results.

\begin{enumerate}[label=(\roman*)]
    \item Although we evolve the ion fractions with an NEI network, the energy equation uses CIE or PIE cooling tables for computational efficiency, following \citet{ji2019simulations}. Turbulent mixing physically brings hot gas to temperatures where radiative losses are efficient, but imposing equilibrium cooling on this gas may also shorten the cooling time. As a partial test, \citet{ji2019simulations} compared the equilibrium cooling curve with the time-dependent isobaric cooling curve of \citet{gnat2007time} and found little change in ion column densities, although the prescribed cooling trajectory does not capture the additional NEI induced by turbulent mixing. Future calculations will couple radiative cooling directly to the evolving ion populations to assess these effects consistently.

    \item Since we primarily focus on emission in the $0.5$--$2$~keV band, our CX calculation includes selected high-charge ions of C, N, O, and Ne, so it does not provide a complete CX reaction network or spectrum. Electron capture by lower-charge ions such as \ion{O}{6}, whose cascades fall outside this band, is omitted. Such reactions can nevertheless alter the ion populations, affecting their absorption and emission signatures and potentially modifying radiative cooling rates. Future extension of the network to lower charge states and other elements will enable joint UV and X-ray diagnostics of the same interfaces.

    \item The CGM photoionization run assumes a spatially uniform metagalactic background and neglects self-shielding, local radiation sources, and radiative transfer. Nevertheless, our runs without photoionization and with the optically thin background represent two useful limiting cases, providing reference points for assessing the sensitivity of CX emission to external illumination. Radiation from nearby stars, an active nucleus, or the cooling interface itself can further change both the donor abundance and the metal charge states, particularly in GC and HVC applications. In addition, the neutral hydrogen abundance is calculated assuming local CIE or PIE. We suspect that the saturated CX surface brightness is not strongly sensitive to this assumption, since the ion-supply argument suggests reduced sensitivity to donor density. Future calculations should improve these approximations.

    \item The simulations describe idealized, initially plane-parallel, single-fluid hydrodynamic mixing layers. Magnetic fields, anisotropic thermal conduction, and physical viscosity may change entrainment, e.g., magnetic tension can suppress the growth of mixing layers \citep{ji2019simulations,zhao2023simulations,das2024magnetic}. In realistic multiphase systems, cold-hot interfaces can be curved and interact with neighboring interfaces, unlike the isolated planar layer considered here, as discussed in \S\ref{subsec:cx_dominance}. Nevertheless, as the first study to include NEI and CX in turbulent mixing layers, our results provide a baseline for understanding the properties and behavior of CX emission at dynamic cold-hot interfaces, and future work will explore these effects in more realistic multiphase systems.
\end{enumerate}

\begin{acknowledgments}
    YZ and SJ are supported by the Natural Science Foundation of China (grants 12522301, 12192223, and  12361161601), the National Key R\&D Program of China (No. 2023YFB3002502), the China Manned Space Program (grants CMS-CSST-2025-A08 and CMS-CSST-2025-A10), and the Xiyuan Undergraduate Research Program at Fudan University. GL is supported by the Natural Science Foundation of China (grant 12373027). SZ is supported by the Natural Science Foundation of China (grant 12273112). ZQ is supported by the China Manned Space Program (grant CMS-CSST-2025-A10). Numerical calculations were run on the CFFF platform of Fudan University and at the Supercomputing Center of Wuhan University.
\end{acknowledgments}

\software{{\small FLASH} \citep{fryxell2000flash},
          {\small Cloudy} \citep{ferland20172017},
          {\small Matplotlib} \citep{hunter2007matplotlib},
          {\small NumPy} \citep{harris2020array},
          {\small SciPy} \citep{virtanen2020scipy},
          {\small yt} \citep{turk2010yt,turk2025introducing},
          {\small ACX2} \citep{foster2020pyatomdb},
          {\small KRONOS} \citep{mullen2016charge}
          }

\bibliography{ref}{}
\bibliographystyle{aasjournalv7}

\end{document}